\documentclass[3p]{elsarticle}
\usepackage{lineno,hyperref}
\modulolinenumbers[5]

\hypersetup{
colorlinks=true,
linkcolor=blue,
filecolor=magenta,      
urlcolor=cyan,
}

\usepackage{comment}
\usepackage{graphicx}
\usepackage{longtable}
\usepackage{float}
\usepackage{booktabs}
\usepackage{amsmath}
\usepackage{amsfonts}
\graphicspath{ {images/} }

\author{Wardah Ali, Eesha Qureshi, Omama Ahmed Farooqi }
\author{Rizwan Ahmed khan\corref{cor1}}
\cortext[cor1]{Corresponding author; Email: Rizwan.Khan@shu.edu.pk;  Rizwan17@gmail.com}
\address{Faculty of IT, Department of Computer Science, Salim Habib University, Karachi, Pakistan.}

\begin{document}
	\begin{frontmatter}
		
\title{Pneumonia Detection in Chest X-Ray Images : Handling Class Imbalance}






\begin{abstract}
People all over the globe are affected by pneumonia but deaths due to it are highest in Sub-Saharan Asia and South Asia. In recent years, the overall incidence and mortality rate of pneumonia regardless of the utilization of effective vaccines and compelling antibiotics has escalated. Thus, pneumonia remains a disease that needs spry prevention and treatment. The widespread prevalence of pneumonia has caused the research community to come up with a framework that helps detect, diagnose and analyze diseases accurately and promptly. One of the major hurdles faced by the Artificial Intelligence (AI) research community is the lack of publicly available datasets for chest diseases, including pneumonia . Secondly, few of the available datasets are highly imbalanced (normal examples are over sampled, while samples with ailment are in severe minority) making the problem even more challenging.  In this article we present a novel framework for the detection of pneumonia. The novelty of the proposed methodology lies in the tackling of class imbalance problem. The Generative Adversarial Network (GAN), specifically a combination of Deep Convolutional Generative Adversarial Network (DCGAN) and Wasserstein GAN gradient penalty (WGAN-GP) was applied on the minority class ``Pneumonia'' for augmentation, whereas Random Under-Sampling (RUS) was done on the majority class ``No Findings'' to deal with the imbalance problem. The ChestX-Ray8 dataset, one of the biggest datasets, is used to validate the performance of the proposed framework. The learning phase is completed using transfer learning on state-of-the-art deep learning models i.e. ResNet-50, Xception, and VGG-16. Results obtained exceed state-of-the-art. 
  
\end{abstract}
\begin{keyword}
\texttt{Medical Imaging \sep Deep Learning \sep Generative Adversarial Network (GAN) \sep Pneumonia \sep Deep Convolutional Generative Adversarial Network (DCGAN)}

\end{keyword}
\end{frontmatter}

\section{Introduction}

Pneumonia is an inflammation of the bronchi, alveoli, bronchioles, and interstitial lungs \cite{ZH2}. The most common types of pneumonia are viral and bacterial pneumonia which possess significant health threats \cite{SMITH20}. Pneumonia substantially occurs due to pathogenic microbial infections, immune function damage, allergies as well as drug factors \cite{HARRIOTT201}.

Pneumonia is a major infectious source of death in children worldwide. In accordance with the statistics issued by the World Health Organization (WHO), 740,180 children that were below the age of 5 died because of pneumonia in 2019 \cite{Pneumonia2021}. People all over the globe are affected by pneumonia but deaths due to it are highest in Sub-Saharan Asia and South Asia. In 2017, more than half of the deaths in five under-developed / developing countries i.e. the Republic of Congo, Pakistan, India, Nigeria, and Ethiopia were from childhood pneumonia \cite{owidpneumonia}. Pneumonia is the third leading cause of death in elderly deaths (people greater than or equal to 80 years of age) in Japan \cite{Kondo2017}.


In recent years the overall incidence and mortality rate of pneumonia, regardless of the utilization of effective vaccines and compelling antibiotics, has escalated. Thus, pneumonia remains a disease that needs spry prevention and treatment. Computer-Aided Diagnosis (CAD) is a very popular technique that assists doctors to detect and interpret various types of abnormalities in medical imaging, to diagnose and analyze diseases accurately and promptly \cite{Doi2007, SHAH20221}. 

Conventionally, manual inspection of chest X-rays is done by the radiologist in order to detect and diagnose pneumonia and other lung diseases, however, it can lead to a prolonged diagnosis process and certain undesirable results. For example,

\begin{enumerate}
	\item About 2/3 of persons around the world still do not have the means to get their disease diagnosed by a radiologist in accordance to a report by the World Health Organization (WHO).
	
	\item Fatigue and the concentration of medical experts/radiologists can affect diagnosis.
	
	\item Inspection of a large number of X-rays on a daily basis can be exhausting and sometimes can lead to wrong diagnosis .
	
	\item Availability of a medical expert at all times is prohibitive.
	
\end{enumerate}

To cater to the above-mentioned drawbacks and challenges of manual inspections of X-rays, CAD systems are leveraging the power of Artificial Intelligence (AI) to detect patterns in the data in order to predict or help medical practitioners in predicting disease \cite{Yu2018}. Generally CAD techniques applied to detect lung infection and diseases involve different imaging modalities, including chest X-rays, Magnetic Resonance Imaging (MRI), chest CT, bronchoscopy, etc \cite{BAGCI201272}. Chest X-ray data is mostly used in the detection of pneumonia as it is cost-effective and does not expose patients to harmful radiations \cite{Nishio2020, gulati2021lung}. 

AI based systems learn patterns from the data (X-rays) and make predictions based on those patterns when new / unseen data (X-ray) is fed. Such AI based CAD systems use machine learning (ML) classifiers for unfolding patterns and making predictions \cite{9794709, hwang2021covid, WANG2021107613}. Generally, AI algorithms can be divided into two sub-categories based on how they unfold patterns in the data: 

\begin{enumerate}
	\item  Conventional machine learning: they need handcrafted features (measurable distinct quantity) for making predictions. 
	
	\item Contemporary machine learning: these latest algorithms process raw data (image in our case), automatically extracts features (learn representations
	from data with multiple levels of abstraction \cite{LeCun2015})  based on mathematical optimization. These algorithms are commonly characterized as Deep Learning (DL)/ Deep Neural Network (DNN) algorithms. 
\end{enumerate}



Generally, machine learning based systems follow three steps to make predictions. First is data or dataset gathering (DL algorithms require much larger dataset as compared to conventional ML algorithms), second step is extraction of discriminative features (discriminative features maximizes inter-class variance and minimizes intra-class variance) and last step is training model / classifier that makes prediction. 

As mentioned above, medical practitioners prefer chest X-ray to detect pneumonia. During literature review, it is also observed that chest X-ray is preferred for machine learning systems as well \cite{Zhang2021, hwang2021covid, WANG2021107613}. This is due to the fact that other imaging modalities including Magnetic Resonance Imaging (MRI) and Computed Tomography (CT) are complex in nature as they capture images in more than two spatial dimensions making the system training complex. Secondly due to cost, recording such data in large quantities is difficult.

Despite the fact that AI is helping to develop effective models for medical image analysis and early diagnosis, there remains quite a number of difficulties / challenges to be resolved. For Example

\begin{enumerate}
	\item Availability of public datasets with large amount of data required for contemporary data hungry models i.e. Deep Neural Network (DNN) \cite{LeCun2015}, Convolutional Neural Network (CNN) \cite{HMRZ},  Recurrent Neural Network (RNN) \cite{LDong18} etc. 
	
	\item Second problem that is associated with available datasets is skewed distribution of data or class imbalance \cite{Gao2020}. Data imbalance occurs whenever one of the classes has more samples than other classes. In our case, we used 'Chest X-Ray8' dataset \cite{wang2017chestx}, one of the most widely used and large publicly available dataset related to lung diseases (refer Section \ref{DS} for discussion on dataset). In this dataset, X-rays with no-findings are around 80\%. In such scenarios, ML classifiers generally focus on correctly classifying majority class while ignoring or misclassifying very important minority class samples. 
	
	\item In the absence of large and balanced datasets, AI / ML scientists use transfer learning approach to get the benefits of state-of-the-art pre-trained DNN models \cite{KHAN201961}. In such cases a very deep (network with many layers) is used even for binary classification problems. This leads to a suboptimal solution, with more than required time and space complexity \cite{STcomplex}. 

\end{enumerate}

The scientific contributions presented in this article are presented below: 

\begin{enumerate}
	\item   As mentioned earlier, we used Chest X-Ray8 dataset  \cite{wang2017chestx}. In this dataset, X-rays with ``no-findings'' are around 80\%. Thus, the data distribution is severely skewed. To deal with this problem we presented an efficient solution. The Generative Adversarial Network (GAN) \cite{goodfellow2014generative}, specifically a combination of Deep Convolutional Generative Adversarial Network (DCGAN) \cite{radford2015unsupervised} and Wasserstein GAN gradient penalty (WGAN-GP) \cite{gulrajani2017improved} was applied on the minority class ``Pneumonia'' for data augmentation, whereas Random Under-Sampling (RUS) was done on the majority class ``no-findings'' to deal with the imbalance problem. Refer to Section \ref{da} for discussion. 
	
	
	\item Secondly, we used transfer learning approach to leverage the state-of-the-art pre-trained Convolutional Neural Networks (CNN) for prediction of Pneumonia.  Refer to Section \ref{tl} for discussion on how transfer learning is applied. Results obtained after dealing with class imbalance problem and application of transfer learning are presented in Section \ref{res}.
	
\end{enumerate}

\section{Literature Review} \label{sota}

By adopting data-driven decisions, various domains are embracing the potential AI and ML techniques have, in-order to boost efficiency \cite{Davenport2020, SHAH20221}. AI's successful contribution includes better quantitative assessment in identifying intricate image patterns from data in a robust and automated manner. Secondly, it is also used as a tool to assist physicians and radiologists for various tasks. 


\subsection {Literature review: CNN based architectures} \label{cnnLit}

Advancement in the hardware for parallel computing i.e. (The graphics processing unit (GPU) \cite{4490127}) and the surge in development of learning algorithms \cite{Hinton2006, Srivastava2014}  made the learning / parameter learning of deep models like deep neural networks(DNN) and Convolutional Neural Networks (CNN) possible. The breakthrough in deep learning \cite{LeCun2015} made the way for unprecedented progress in Computer Vision i.e detection, classification and semantic segmentation \cite{Girshick2014, Voulodimos2018}.  CNNs are the foundation for the majority of Deep Learning (DL) techniques that have produced state-of-the-art results on visual inputs, like images and movies \cite{lecun1998gradient, Fukushima1980}. Convolutional Neural Networks, explored in the early 90's \cite{Fukushima1980} have grown in popularity as a machine learning technique for a variety of applications, including medical image analysis \cite{SHAH20221}.

After the advent of DNN and CNN, most of the work that utilizes robustness of AI and ML for medical image analysis is based on CNN, a deep neural network architecture adopted for analyzing visual stimuli. The reason for the adoption of CNN in image analysis tasks is its superior performance in robustly learning meaningful features from visual stimuli and thus improving state-of-art-results for classification. 

One of the groundbreaking work done that leverages CNN for the analysis of lung diseases is done by Rajpurkar et al. \cite{arxiv17110}. They proposed 121-layer CNN, called CheXNet. The CheXNet not only predicts lung disease, if any, but also produces heat map at the output to indicate the region of interest in the X-ray that played a major role in the prediction. They used chest x-ray image dataset of NIH labeled with 14 diseases, called ChestX-ray14 \cite{wang2017chestx}.ChestX-ray14 is an extension of ChestX-ray8 dataset. Rajpurkar et al. reported that CheXNet obtained  an F1 score of 0.435, higher than the radiologist average of 0.387.

In another research \cite{Aplos}, the state-of-the-art CNN architectures i.e. VGG16/19, Xception, Inception and ResNet were utilized for pneumothorax’s detection using NIH ChestX-ray14 data. Pneumothorax is a complication induced by pneumonia \cite{EKANEM2021437}. In this research authors extracted 13292 frontal chest X-rays. In the transfer learning setting, they combined multiple state-of-the-art CNN architecture's final feature maps. The proposed framework obtained AUC of 0.75.

Ozturk et al. \cite{Ozturk2020} introduce the Darknet model as a new method for automatic COVID-19 detection using unprocessed chest X-ray images. Their proposed methodology is made to provide accurate diagnostics for binary classification tasks (COVID vs. No-Findings) and multi-class classification tasks (COVID vs. No-Findings vs. Pneumonia). They reported a multi-class classification accuracy of 87 percent and a binary classification accuracy of 98.08 percent on 25 COVID-19, 100 normal, and 100 pneumonia images.

AlMamlook et al. \cite{AlMamlook2020} proposed a model to increase accuracy and efficiency for the classification of normal(healthy) from abnormal (sick) Chest X-rays. They used seven  state-of-the-art machine learning models and techniques \cite{KHAN20131159} along with well-known Convolution Neural Network (CNN) models achieving an overall accuracy of 98.46\%. These include Random Forest (RF), Decision Tree (DT), Naive Bayes (NB), Support Vector Machine (SVM), Linear Discriminant Analysis (LDA), K-Nearest Neighbour (KNN), and Logistic Regression (LR).

Alhudhaif el at. \cite{ALHUDHAIF2021115141} proposed CNN (DenseNet-201 \cite{Huang2017}) based architecture for detection of COVID-19 pneumonia. They used 1,218 chest X-ray images dataset, collected from publicly available databases. The dataset had only 368 COVID-19 pneumonia X-rays. Validation was performed using 5-fold cross-validation.  Accuracy, precision, recall, and F1-scores of 94.96\%, 89.74\%, 94.59\%, and 92.11\%, respectively were reported.

Nikolaou et al. \cite{Nikolaou2021} used COVID-19 Radiography database \cite{DatabaseKag} with 15,153 X-ray images. Their proposed novel framework was based on the transfer learning approach. They extended pre-trained EfficientNetB0 \cite{pmlr-v97-tan19a} with a dense layer of 32 neurons on top to detect COVID-19 pneumonia. Conventional data augmentation techniques i.e. rotation, image clipping, random zoom etc; were used to cater problems that arises from using small amounts of data with class imbalance. Authors have reported 95\% accuracy for their proposed model.

Das et al. \cite{Das2021} applied ensembling technique on the predictions obtained from state-of-the-art Deep Convolutional Neural Networks (DCNN) which include DenseNet201 \cite{Huang2017}, Resnet50V2 \cite{He2016} and Inceptionv3 \cite{szegedy2015going}. Each network was trained individually on X-ray collected from different open source public repositories. The dataset includes 538 X-rays of COVID +ve patients and 468 X-rays of COVID –ve patients.  Das et al. reported classification accuracy of 91.62\%.

Singh et al. \cite{Singh2022} proposed to use Quaternion Convolution neural network (QCNN) \cite{Zhu_2018_ECCV} which is an extension of CNN. QNNs are better in understanding and analyzing relationships between color channels of RGB image and thus have superior performance in extracting features tangled in the color hierarchies. Their proposed framework was tested on the Chest X-Ray dataset that has only 5,863 X-Ray images. They reported accuracy of 93.75\% for Pneumonia detection.

Gour and Jain \cite{GOUR202227} used stacked / ensemble convolutional neural network model for detection of  COVID-19 pneumonia from the chest X-ray and CT images. Gour and Jain applied transfer learning approach on pre-trained VGG19 \cite{Simonyan2015} and the Xception \cite{chollet2017xception} models. Then, they efficiently combined the prediction of these models. The model was tested on 3,040 chest X-ray images (546 pneumonia +ve images) dataset. The X-ray images in the dataset were gathered from three publicly available datasets. Authors reported sensitivity of 97.62\%.

Szepesi et al. \cite{SZEPESI2022} proposed novel CNN architecture, inspired from VGG-16 architecture \cite{Simonyan2015}. The novel architecture carefully places the dropout layer (mainly used to prevent over-fitting \cite{GoodfellowB2016}) in the convolution part of the network. They used the same dataset of 5,863 X-Ray images as used in the study by Singh et al. \cite{Singh2022}, mentioned earlier. Proposed architecture achieved  97.2\% accuracy for Pneumonia detection.

\subsection {Literature review: Generative Adversarial Networks (GANs) application to deal with class imbalance}

Architectures / frameworks that have been mentioned in the earlier subsection (Section \ref{cnnLit}) were tested on small and imbalanced datasets (except few) and didn't explicitly deal with class imbalance problem. There were only a few frameworks that tried to deal with the class imbalance problem using traditional method of data augmentation. They created more samples of minority class by applying geometric transformations to original images, e.g. rotations, zooming, mirroring etc. These transformations are known to change or disrupt geometric / orientations based features present in the data \cite{mariani2018bagan}.

Sometimes gathering / recording large labeled dataset in clinical settings is often difficult especially when the number of patients with a certain medical condition are not enough \cite{Borjali2020}. Secondly, medical practitioners tend to recommend some medical tests i.e. X-ray, in order to reject the possibility of some diseases. Thus, the number of observations in the dataset with normal conditions outnumber the number of observations with detected disease \cite{wang2017chestx}. This leads to class imbalance problem where the number of samples belonging to different classes are not balanced or the distribution of data is skewed. Data imbalance problem significantly degrades performance of machine learning models \cite{Leevy2018}. Generally, if the dataset is imbalanced, the learning algorithm or the objective function develops bias towards the majority class. Thus, a very important minority class pattern is not catered efficiently.

The Generative Adversarial Networks (GANs) \cite{goodfellow2014generative} have attained success in computer vision and natural language processing as being amongst one of the most innovative deep learning models in recent years. The concept of game theory \cite{roughgarden2010algorithmic} is used by adversarial networks or GANs in particular, as they are trained to play a minimax game \cite{wang2017irgan} with a discriminator and a generator network, which aims to maximize a given objective function whereas a discriminator attempts to minimize the very same objective function, thus the term ``adversarial''. The basic idea of using  ``generator'' and a ``discriminator' is to achieve training indirectly. Generator keeps on creating data samples that resemble real samples by learning features of different classes, while the discriminator classifies whether a sample is real or fake. After extensive training synthetically generated samples begin to resemble the real samples.

Across multiple data regimens, GAN data-augmented models and standard augmented models trained on Chest X-Ray images (CheXpert dataset \cite{irvin2019chexpert}) were compared by Sundaram et al. \cite{Sundaram2021}, their findings demonstrate that GAN-based augmentation proved to be a useful method for addressing medical datasets with class imbalance problems. They further indicated their comparison results through AUC performance gains.

A GAN-based framework was proposed by Malygina et al. \cite{Malygina2019} in order to cater to the class imbalance problem. They used CXR14 dataset of chest X-rays \cite{wang2017chestx}, that has 84312 samples of ``normal'', only 9838 samples of ``pneumonia'', 10963 samples of ``fibrosis'' and 10963 samples of ``pleural-thickening''. Thus, the dataset is highly imbalanced. They applied CycleGAN and trained it on unpaired images such that it generates images from the opposite class for each input image. Furthermore, their results show that the classifier performance greatly improved for the pneumonia class however they couldn't achieve considerable changes for pleural-thickening, and also observed degradation of classifier quality on fibrosis, thus they concluded that the proposed GAN architecture is insufficient to handle such complex instances as fibrosis. 

The authors of \cite{Srivastav2021} used a transfer learning based approach by utilizing the VGG-16 model for pneumonia detection. The dataset used in their model was the Mendeley data of chest X-rays \cite{kermany2018large}. The dataset has 5856 images of chest X-rays.  To cater dataset imbalance problem, they used 
Deep Convolutional Generative Adversarial Network (DCGAN) augmentation technique \cite{radford2015unsupervised}. DCGAN augmented X-ray images of minority class. Authors reported accuracy of 94.5\% on binary classification task of predicting ``Pneumonia'' and ``no finding''.

Sundaram and Hulkund \cite{sundaram2021gan} analyzed and concluded positive impact of GAN based data augmentation technique on the efficacy of DNNs in diagnosing lung diseases from chest X-rays. In their study CheXpert dataset \cite{irvin2019chexpert} was used. 

While posing the problem from a different perspective,  Luyi et al. \cite{han2022gan} proposed to use ``Rib Suppression GAN'', or RSGAN in order to robustly detect pulmonary diseases. As per this article, suppressing rib from the chest X-ray has a positive impact on detection accuracy. The authors employed adaptive loss to suppress rib residue and preserve other details in the X-ray images. The residual map was used to characterize the difference of intensities between Chest X-ray and the corresponding rib-suppressed data. Dataset from two publicly available CT datasets and four chest X-ray datasets was used. Luyi et al. showed that by combining chest X-ray with its corresponding rib suppressed image can help in achieving better accuracy then using either of these two image modalities separately.

Shah et al. \cite{shah2022dc} used Deep Convolutional Generative Adversarial Network (DCGAN) \cite{radford2015unsupervised} to cater the problem of class imbalance and shortage of data samples for all the classes i.e. normal, pneumonia and COVID-19. Synthetically generated images / data was validated using $k$-mean clustering technique \cite{1056489}. Only those synthetically generated samples / data points were retained in the dataset that were classified in the correct clusters. The validated dataset was then fed to The EfficientNetB4 \cite{pmlr-v97-tan19a}, a convolutional neural network (CNN) architecture,  for training. The reported experiments achieved promising results of 95\% area under the curve (AUC).



\section{Dataset} \label{DS}

\begin{figure}[H]
	\centering
	\includegraphics [scale=0.9]{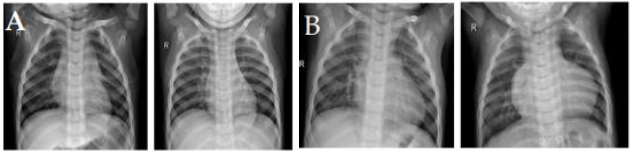}
	\caption{a) No findings / normal X-ray b) Pneumonia +ve X-ray}
	\label{fig:PvsNF}
\end{figure}

The Chest X-Ray8 (CX8) dataset \cite{wang2017chestx} was used in this research study. This dataset is one of the most extensive chest X-ray images dataset publicly available for research purposes. The dataset contains 112,120 frontal view X-ray images of 32,717 unique patients. The X-rays are extracted from the DICOM (Digital Imaging and Communications in Medicine \cite{mildenberger2002}) file. Then files were resized to 1024 $\times$ 1024 spatial resolution.

\begin{figure}[!htb]
 	\centering
 	\includegraphics [scale=0.7]{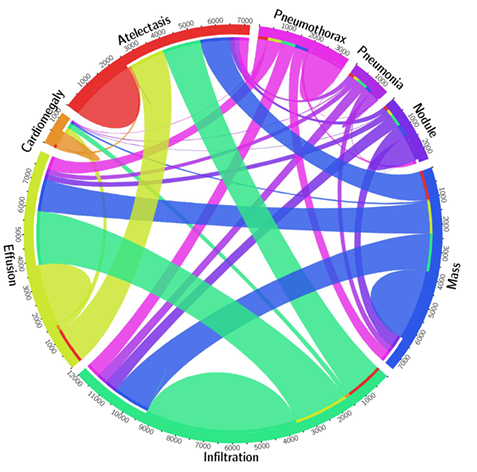}
 	\caption{The proportions of X-rays with multi-labels and the labels’ co-occurrence statistics \cite{wang2017chestx}}
 	\label{fig:multilabel}
 \end{figure}

Each X-ray in CX8 could have multiple labels. Total of eight diseases or labels (atelectasis, cardiomegaly, effusion, infiltration, mass, nodule, pneumonia, and pneumothorax) are present in the dataset, refer to Figure \ref{fig:multilabel} to see multi-label statistics. The labels are extracted / text mined by analyzing corresponding radiological reports with the help of natural language processing (NLP) \cite{collobert2011natural} tools i.e. DNorm \cite{leaman2015challenges} and MetaMap \cite{aronson2010overview}.  As the labels are automatically extracted using NLP the probability of error in labeling exists, specially when the samples have multiple labels. 

In this research study we focused on binary classification problem e.g. to detect whether a given X-ray is normal or is  pneumonia positive. There is a total of 60,0000 X-ray images with label no finding that means those images have not been diagnosed with any of those above mentioned diseases. While, only about 1\% of X-ray images are labeled as having pneumonia. So, there is a strong class imbalance in the dataset. 

 
\section{Data Augmentation: Dealing with class imbalance problem} \label{da}

As mentioned above, in this study we used ``Chest X-Ray8'' dataset \cite{wang2017chestx}, one of the most widely used and large publicly available dataset related to lung diseases. In this dataset, X-rays with ``no-findings'' are around 80\%. Thus, the data distribution is severely skewed. To deal with the class imbalance problem we propose an efficient solution. We propose to use Generative Adversarial Network (GAN) \cite{goodfellow2014generative}, specifically a combination of Deep Convolutional Generative Adversarial Network (DCGAN) \cite{radford2015unsupervised} and Wasserstein GAN gradient penalty (WGAN-GP) \cite{gulrajani2017improved}. The propose method generates sample for minority class i.e. ``Pneumonia''. Whereas Random Under-Sampling (RUS) \cite{khushi2021comparative} was done on the majority class i.e. ``no-findings''.



\begin{figure}[!htb]
	\centering
	\includegraphics [scale=0.65]{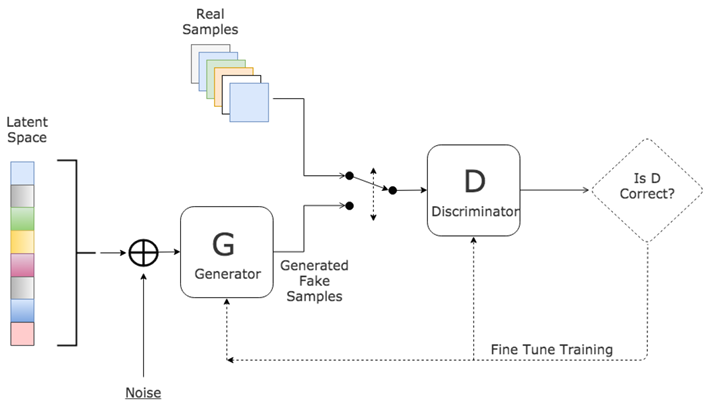}
	\caption{The structure of a Generative Adversarial Network (GAN) \cite{hitawala2018comparative}}
	\label{ganArch}
\end{figure}

GANs create images / data samples that closely resemble the distribution of the original dataset's feature distribution. A generator and a discriminator are the two models in the GAN, simultaneously trained via an adversarial process. The discriminator learns to distinguish  between actual and fake images, while the generator learns to produce images that resemble real images. Until the discriminator is unable to distinguish between actual and fake images, we kept training both of these models. GANs are proved be beneficial in generating hyper-realistic human faces \cite{hu2021exposing, Yang2019}, medical image analysis \cite{kazeminia2020gans}, dealing with class imbalance problem \cite{shah2022dc, han2022gan} and many other applications. Refer to Figure \ref{ganArch} to see architecture of GAN.

The objective function of GAN is given by:

\begin{equation}
	\min_{G} \max_{D}  \mathbb{E}_{x \sim P_r} [log(D(x))] + \mathbb{E}_{\tilde{x} \sim P_g} [log(1-D(\tilde{x}))] 
	\label{eq:gan}
\end{equation}

where $P_r$ is the real data distribution, and $P_g$ is the fake  /generated distribution. $\tilde{x} \sim G(z)$ is used in defining the fake distribution and $z \sim p(z)$ comes from some distribution p.

The discriminator loss is defined by: 
\begin{equation}
J_D = -\mathbb{E}_{x\sim P_r} [log(D(x))] - \mathbb{E}_{\tilde{x}\sim P_g} [log(1 - D(\tilde{x}))]
\label{eq:dloss}
\end{equation}

And the generator loss is:
\begin{equation}
	J_G = -J_D
	\label{eq:loss}
\end{equation}

Goodfellow et al. \cite{goodfellow2014generative} in their ground-breaking work demonstrated that in GAN / min-max game learning is  equivalent to minimizing the Jensen-Shannon(JS) divergence \cite{sutter2020multimodal} between the real and fake distribution.

\begin{equation}
\label{eq:gloss}
J_G = -\mathbb{E}_{\tilde{x} \sim p_g}  [log(D(\tilde{x}))]
\end{equation}



Further in 2016, Goodfellow \cite{goodfellow2016nips} showed that training GANs with the above objective function; refer to Equations \ref{eq:gan}, \ref{eq:dloss},  \ref{eq:loss} and  \ref{eq:gloss}; is unstable and non-convergent.

\subsection{Deep Convolutional Generative Adversarial Network (DCGAN)} \label{sec:dcgan}

In this research work we have used the Deep Convolutional Generative Adversarial Network (DCGAN) \cite{radford2015unsupervised} which is an extension of the Generative Adversarial Network (GAN). DCGAN is one of the most well-known and effective GAN implementations for visual stimuli. DCGAN  primarily consists of convolution layers in place of multi-layer perceptron, the convolution layers are implemented without any fully connected or max pooling layers that are used in vanilla GAN. As earlier GAN research exploited momentum to speed up training, Adam \cite{kingma2014adam} extension of stochastic gradient descent with a learning rate of 0.001, was used for optimization for both the generator and discriminator in our model. However, it was empirically determined that in this study learning rate of 0.0002 gives better results.

\subsection{Wasserstein GAN gradient penalty (WGAN-GP)}

As mentioned above, training GAN with loss functions presented in Equations \ref{eq:gan}, \ref{eq:dloss},  \ref{eq:loss} and  \ref{eq:gloss}; is unstable and non-convergent. To overcome this issue we have used  Wasserstein GAN gradient penalty (WGAN-GP) \cite{gulrajani2017improved}. The Wasserstein GAN gradient penalty (WGAN-GP) is a generative adversarial network that uses the gradient norm penalty and the Wasserstein loss formulation to achieve Lipschitz continuity \cite{gulrajani2017improved}. 

Arjovsky et al. \cite{arjovsky2017wasserstein} showed that the JS divergence, along with other common distances, do not provide desired gradient value for training generator. Instead they proposed to use Wasserstein distance to measure the difference between two distributions. Thus, the benefit of The Wasserstein GAN gradient penalty (WGAN-GP) is its convergence. It improves training stability, hence making it easier to train. The details of the Wasserstein loss with gradient penalty are as follows:

\begin{itemize}
	\item The difference between the desired value of the discriminator's output for actual images and the discriminator's output expected value for artificially generated fake images is what constitutes Wasserstein's loss.
	
	\item The discriminator (called the ``critic'' in the original article) aims to increase the gap (refer previous point), whereas the generator aims to decrease it. Equations of discriminator loss and generator are: 
	
	\begin{equation}
		\label{wloss1}
		J_D = \mathbb{E}_{\tilde{x} \sim p_g}[D(\tilde{x})] -  \mathbb{E}_{x \sim p_r}[D(x)]
	\end{equation}

	\begin{equation}
		\label{wloss2}
		J_G = - \mathbb{E}_{\tilde{x} \sim p_g}[D(\tilde{x})]
	\end{equation}

\item WGAN-GP \cite{gulrajani2017improved} employs gradient penalty rather than weight clipping to impose the Lipschitz constraint.
	
	\begin{equation}
		\label{wloss3}
		Loss= \mathbb{E}_{\tilde{x} \sim P_g}[D(\tilde{x})] - \mathbb{E}_{x \sim P_r} [D(x)]  + \underbrace{\lambda \mathbb{E}_{\hat{x} \sim P_{\hat{x}}} [(|| \nabla_x D(\hat{x})||_2 - 1)^2]}_{\text{Gradient Penalty}}
	\end{equation}

where $P_{\hat{x}}$ is sampling distribution of samples along straight lines between
pairs of points sampled from the data distribution $P_r$ and the generator distribution $P_g$. In this research we have empirically found a value of $\lambda$ = 10 to work well. This is the same value as found by  Gulrajani et al. \cite{gulrajani2017improved}.

	\item Batch normalization is not anymore employed in the critic (discriminator) since batch normalization transforms discriminator's ai from mapping  a single input to a single output to mapping from an entire batch of inputs to a batch of outputs \cite{salimans2016improved}. What we require is to be able to determine the gradients of each output relative to its corresponding inputs.


	\item So, the generator's ultimate goal is to raise the mean of the fake output produced by the discriminator. Whereas, the discriminator's objective is Wasserstein loss along with weighted penalty.

\end{itemize}

\begin{figure}[!htb]
	\centering 
	\includegraphics [scale=0.32]{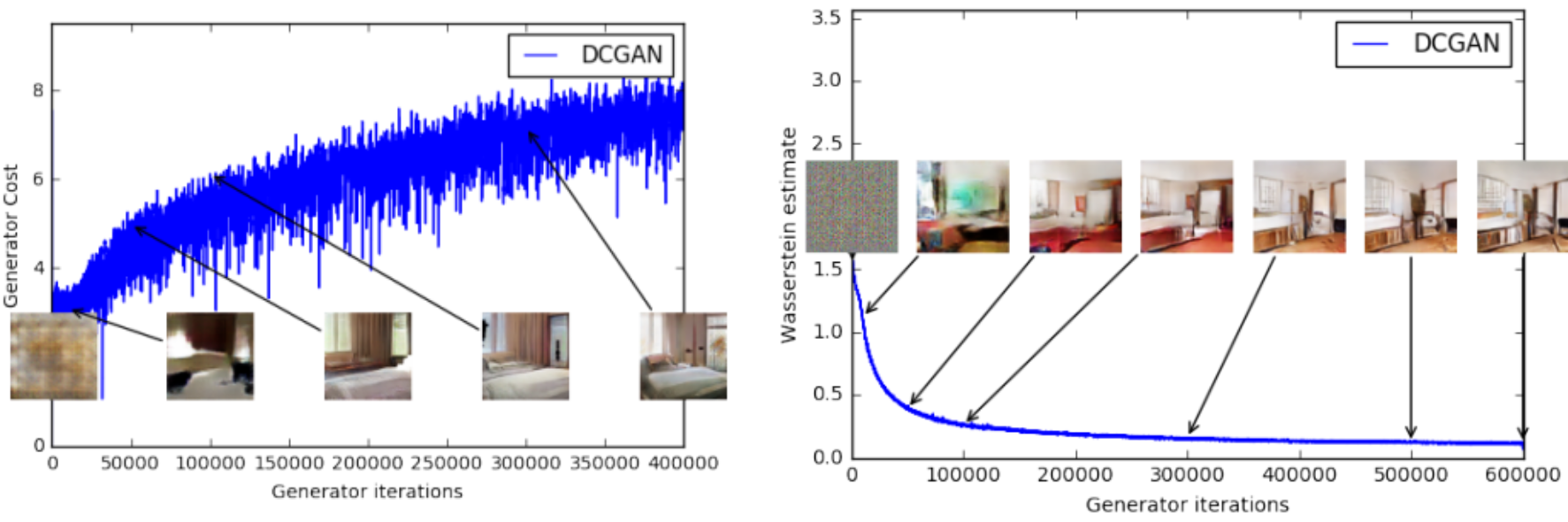}
	\caption{Generator's cost during GAN training (left). Generator's cost during DCGAN training with Wasserstein's loss (right) \cite{arjovsky2017wasserstein}}
	\label{fig:costwgan}
\end{figure}

\begin{figure}[!htb]
	\centering
	\includegraphics [scale=0.7]{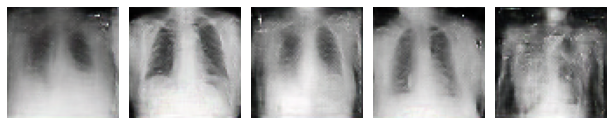}
	\caption{WGAN-GP generated results applied on the minority class ``Pneumonia'' with 64x64 image resolution, these results were further used in our classification task.}
	\label{fig:goodres}
\end{figure}

In legacy GAN, the loss evaluates how effectively it deceives the discriminator instead of measuring the image quality. The generator loss in GAN is not reduced even as the image quality rises as shown in Figure \ref{fig:costwgan} and as a result, we are unable to determine progress from its value. On the other hand, the more desirable image quality is reflected by the WGAN loss function as the loss drops significantly, and generated sample quality also improves as shown in Figure \ref{fig:goodres}. 



We utilized the Wasserstein's loss instead of the Binary Cross Entropy (BCE) loss because of the issues discussed previously. During WGAN training, the discriminator is trained several times for each step, whereas the generator is trained once per step. As a result, we trained the discriminator more than the generator, therefore the generator was updated after every 5 epochs, while the discriminator was updated after each epoch. We applied the convolution layers used in DCGAN in both generator and discriminator. The RMSProp (Root Mean Squared Propagation, an extension to the gradient descent optimization algorithm) optimizer was used for the discriminator while Adam optimizer was used for generator. This setup follows the recommendation of \cite{gulrajani2017improved}. The activation function used in the discriminator was removed as WGAN-GP takes in the difference between the discriminator's output's expected value for real images and its expected value for the generated fake images.



With the help of the method described above the problem of class imbalance was tackled. The samples for  minority class,  i.e. ``Pneumonia'', were generated. Whereas Random Under-Sampling (RUS) \cite{khushi2021comparative} was done on the majority class i.e. ``no-findings''.  The dataset that was generated had following distribution, refer Table \ref{table:imbalance}.

\begin{table}[!htb] 
	\caption{The ChestX-Ray8 dataset after catering the imbalance problem} 
	\label{table:imbalance}
	\centering 
	\begin{tabular}{p{0.2\textwidth}p{0.15\textwidth}p{0.15\textwidth}}
		\hline\hline 
		Class & Samples before augmentation & Samples after augmentation \\ [0.5ex] 
		\hline 
		No Findings & 63,115 & 30,000 \\ 
		Pneumonia & 322 &  30,000 \\ 
		[1ex]
		\hline 
	\end{tabular}
\end{table}

\section{Proposed Framework for Pneumonia Detection} \label{tl}

\begin{figure}[!htb]
	\centering
	\includegraphics [scale=0.2]{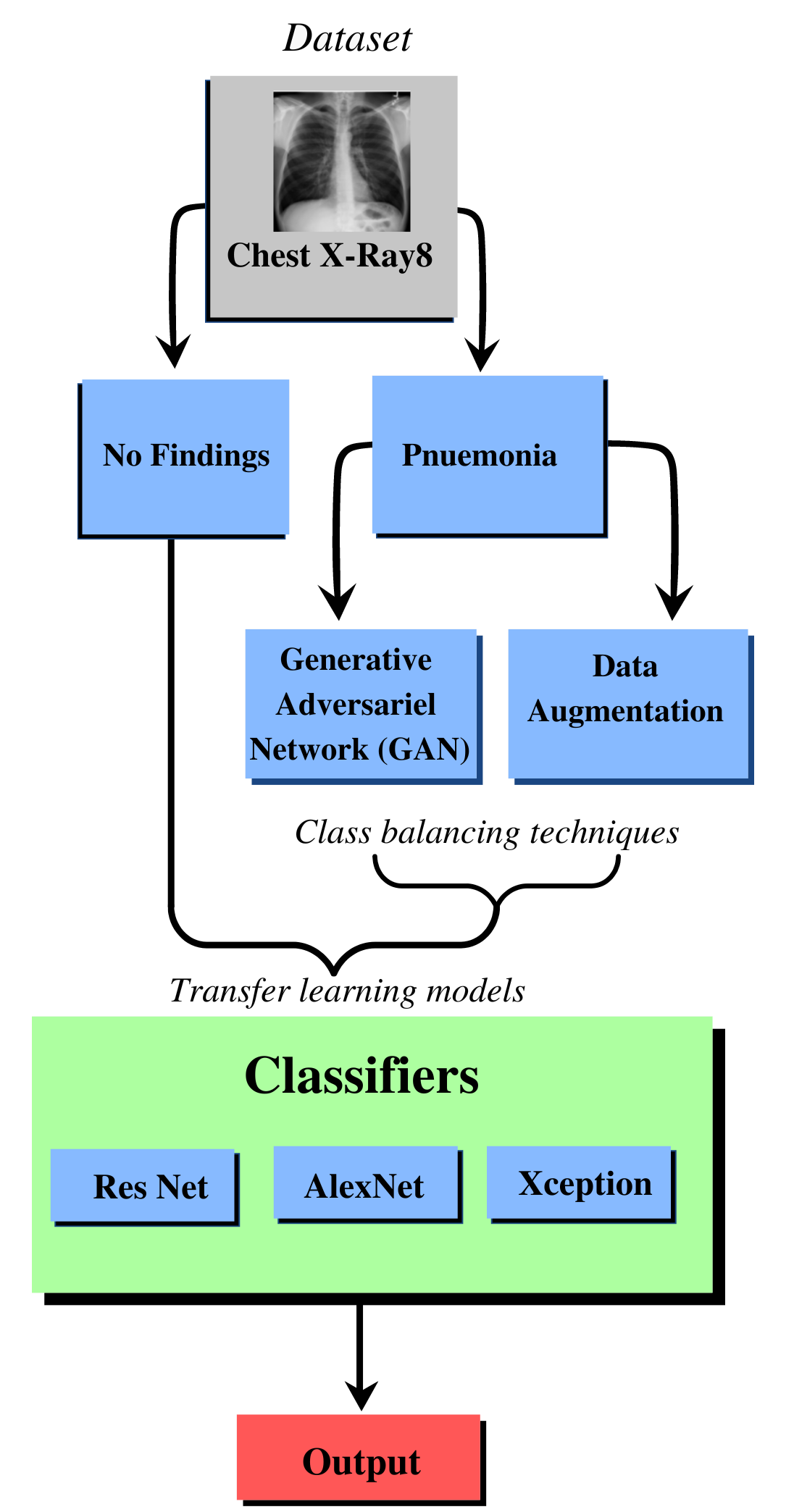}
	\caption{Overview of proposed framework for Pneumonia detection that deals with class imbalance problem}
	\label{figOver}
\end{figure}

The overview of the novel framework that analyzes chest X-ray and deals with class imbalance problem is presented in Figure \ref{figOver}.  The system uses Deep Convolutional Generative Adversarial Network (DCGAN) \cite{radford2015unsupervised} in combination with Wasserstein GAN gradient penalty (WGAN-GP) \cite{gulrajani2017improved} to create realistic samples of minority class. Whereas Random Under-Sampling (RUS) \cite{khushi2021comparative} was done on the majority class. These steps ensured a balanced distribution of both the classes i.e. pneumonia and normal. Balance dataset distribution helps in learning and untangling intricate features from the dataset. 

Proposed novel technique for dealing with class imbalance problem is discussed earlier in Section \ref{da}. This section presents Convolutional Neural Network (ConvNet / CNN) architectures utilized for the classification task. Different state-of-the-art architectures are used to evaluate robustness and appropriateness of synthetically created data point / X-rays.

Two methods—Denovo and Transfer Learning (TL) are used to train CNN models. Denovo, as its name implies, creates a full CNN architecture from scratch, allowing it to learn features from the dataset in the most efficient manner possible. The problem with this approach is that it needs extensive compute power and a large dataset to train the network from scratch.  

Transfer Learning (TL) is the second method for training the CNN architecture. Transfer learning allows to have different distribution of training and test samples \cite{Pan2010}. Basic idea of TL is re-usability of trained mode. In order to adapt a pre-trained CNN model to a specific problem and dataset, the TL approach enables re-training of only a limited number of layers (often the last layers). TL approach mitigates bottleneck of availability of extensive compute power and large dataset \cite{Liu1999, KHAN201961}.


Typically, state-of-the art pre-trained CNN architectures from ImageNet-Large-Scale Visual Recognition Challenge (ILSVRC) \cite{Russakovsky2015} are utilized for TL. Some of the most used  state-of-the art pre-trained CNN architectures are AlexNet \cite{Krizhevsky2012}, VGG \cite{Simonyan2015}, ResNet \cite{He2015}, DenseNet \cite{huang2017densely}, Xception \cite{chollet2017xception} etc. In this study, we employed ResNet, VGG and Xception  to verify the robustness of technique proposed to mitigate the issue of class imbalance. Brief description of these CNN architectures is given below.

\subsection{VGG16}
\begin{figure}[!htb]
	\centering
	\includegraphics [scale=0.38]{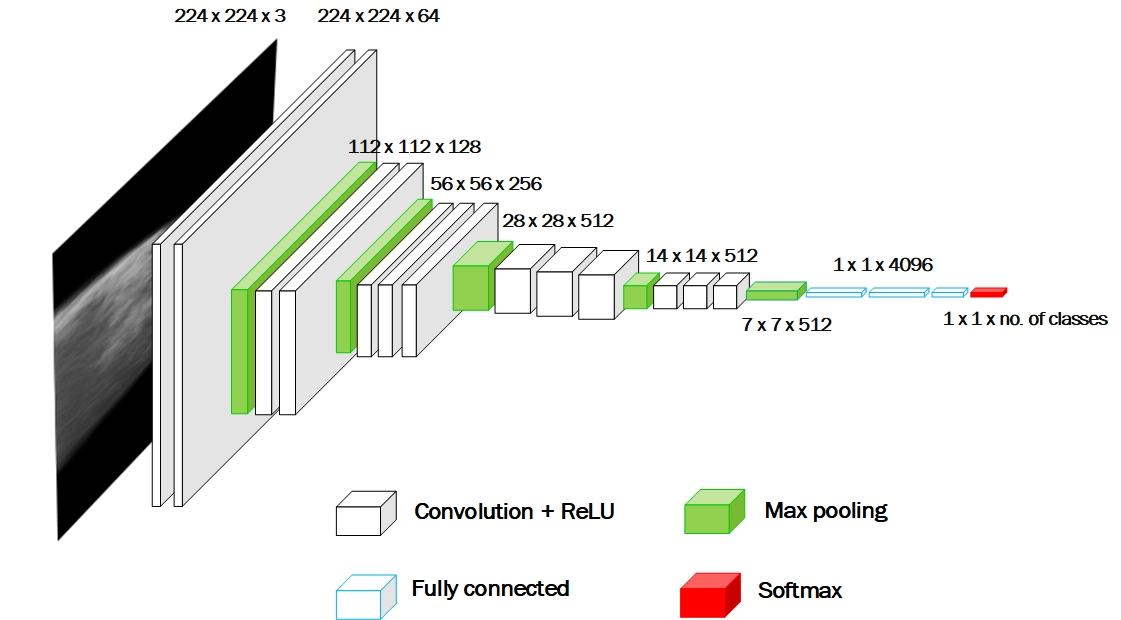}
	\caption{Architecture of VGG-16: Convolution, max-pooling, and dense layers by \cite{sajid2022breast}}
	\label{vggFig}
\end{figure}

VGG-16 was proposed in 2014 by Andrew Zisserman and Karen Simonyan of the Visual Geometry Group (VGG) Lab at Oxford University \cite{Simonyan2015}. VGG was the best performing architecture ILSVRC 2014. Apart from its performance, VGG's uniformity in architecture (refer to Figure \ref{vggFig}) is another factor for its appeal. VGG-16 has sixteen weighted layers (learnable parameters). There are 21 layers altogether: 13 convolutional layers, 3 dense layers, and 5 max pooling layers. Only 16 of its layers have learnable parameters, hence the name VGG-16. Spatial dimension of visual stimuli input to VGG is 224 $\times$ 224. Rectified Linear Unit (ReLU) is used as an activation function, with the Softmax classifier at the last
layer.

\subsection{Residual Networks (ResNet50)}
In theory, the deeper the neural network the better the performance of the network as the intuition is that by adding additional layers the network can progressively learn more complex features. The issue with deeper networks is difficulty in training (vanishing gradients) and degradation in accuracy. To answer this issue, He et al. \cite{He2015} propose ResNet.  

\begin{figure}[!htb]
	\centering
	\includegraphics [scale=0.7]{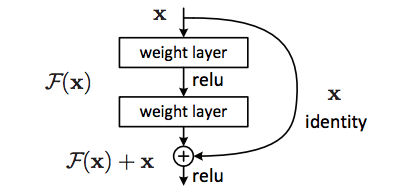}
	\caption{Residual Learning: a building block (Kaiming He et al. \cite{He2015})}
	\label{resnet}
\end{figure}

ResNets resolved the problem of training very deep networks with the introduction of residual blocks with skip connections. As the name suggests, skip connection passes activation value of a  layer to further layers by skipping layers in between, refer to Figure \ref{resnet}. Thus through these skip connections (skipping layer that degrades performance), ResNet solves the degradation problem, which allows it to take advantage of adding hundreds of new layers without diminishing results. With these residual blocks, architecture fits residual mapping: 

\begin{equation}
	y = F(x,{W_i}) + W_s x	
\end{equation}

\subsection{Xception}

Xception or Extreme inception \cite{chollet2017xception} is based on stronger hypothesis of the inception module i.e cross-channel correlations and spatial correlations can be mapped entirely separately \cite{chollet2017xception}. The Inception architecture \cite{szegedy2015going} is based upon the idea how a convolutional network can approximate and cover easily available dense components through its optimal local sparse structure. The Inception performs 1 $\times$ 1, 3 $\times$ 3, and 5 $\times$ 5 convolution computations within the exact same layer in the network after stacking the output of all these filters along the channel dimension and moving on to the following layer. Xception is an extension of the inception module, following depth wise separable convolution. It has 36 convolutional layers with residual connections. All convolution and separable convolution layers are followed by batch normalization. The residual connections proposed by He et al. \cite{He2015} helps the architecture in convergence.


\subsection{Results and Discussions} \label{res}

With the help of the method described in the Section \ref{da} the problem of class imbalance was tackled. The samples for  minority class,  i.e. ``Pneumonia'', were generated. Whereas Random Under-Sampling (RUS) \cite{khushi2021comparative} was done on the majority class i.e. ``no-findings''.  The distribution of the generated dataset is presented in Table \ref{table:imbalance}. The dataset was split in training, test and validation.

\begin{figure}[htb]
	\centering
	\includegraphics [scale=0.7]{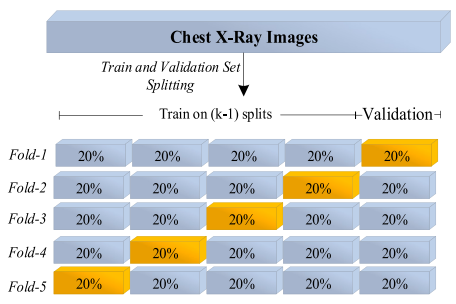}
	\caption{Schematic representation of 5-fold cross-validation procedure \cite{Ozturk2020}}
	\label{figFold}
\end{figure}

80\% of the chest X-ray (CXR) images were used for training. We performed 5-fold cross validation to calculate accuracy of different CNN architectures.  In each of five iteration, 20\% is used to test the accuracy. The process is shown in Figure \ref{figFold}. The accuracy calculated using $k-$fold cross validation is unbiased estimate of generalization accuracy.

As mentioned earlier we used transfer learning approach wit three state-of-art CNN models (hyper-parameters used for training; loss = binary cross entropy, optimizer = Adam and learning rate = 0.001), including

\begin{enumerate}
	\item ResNet50 
	\item VGG16
	\item Xception
\end{enumerate}

\begin{figure}[!htb]
	\centering
	\includegraphics [scale=0.3]{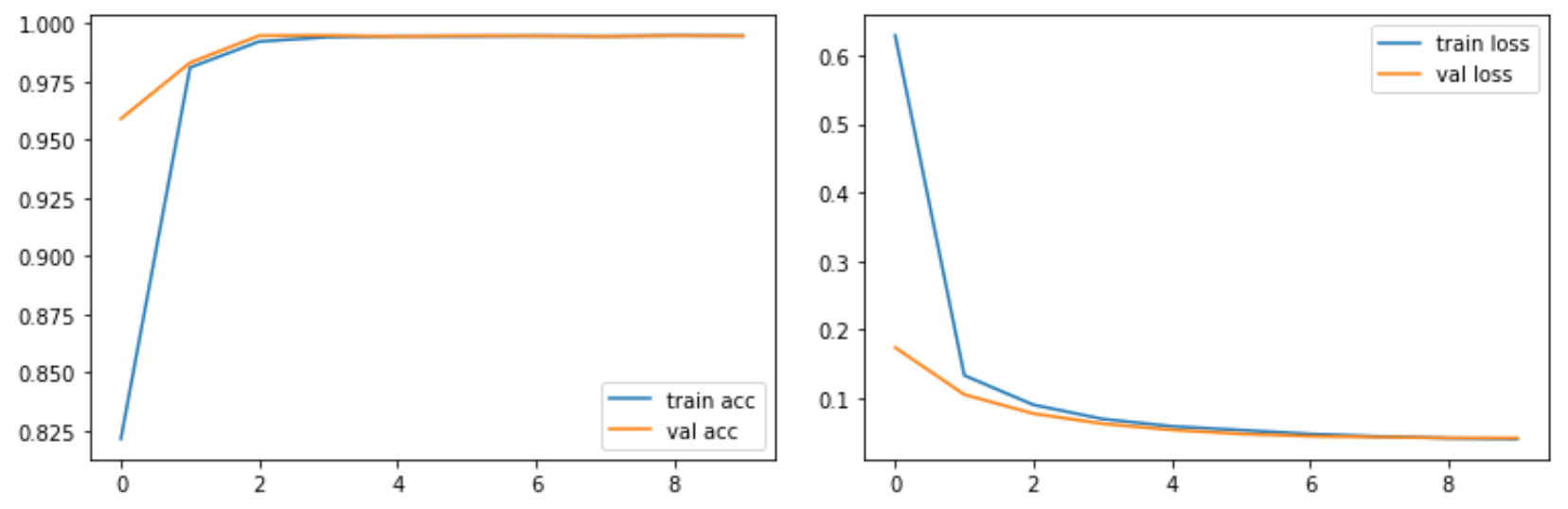}
	\caption{ResNet  model learning: (Left) Training accuracy vs Validation accuracy (Right) Training loss vs Validation loss}
	\label{fig:res}
\end{figure}

\begin{figure}[!htb]
	\centering
	\includegraphics [scale=0.3]{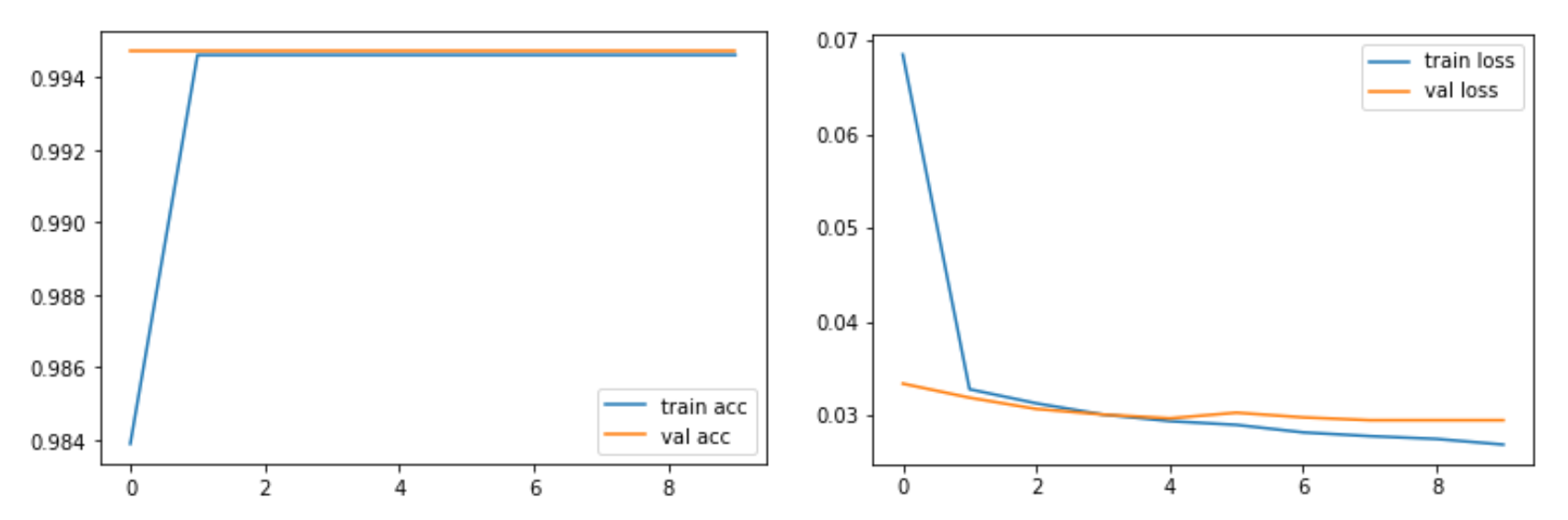}
	\caption{VGG model learning: (Left) Training accuracy vs Validation accuracy (Right) Training loss vs Validation loss}
	\label{fig:vgg}
\end{figure}

 The results achieved by the proposed approach are presented in Table \ref{table:res}. Learning curves of two CNN models are presented in Figure \ref{fig:res} and \ref{fig:vgg} for example. 

\begin{table}[!htb]
	\caption{Accuracy of different CNN models} 
	\label{table:res}
	\centering 
	
	\begin{tabular}{c c c c c c} 
		\hline\hline 
		Model & Epochs & Loss & Accuracy & Validation loss & Validation accuracy \\ [0.5ex] 
		\hline 
		ResNet50 & 20 &  47.90\% & 77.96\% & 36.25\% & 89.14\%\\ 
		VGG16 & 20 &  9.06\% & 97.64\% & 8.52\% & 97.86\%\\ 
		Xception & 20 & 6.48\% & 99.18\% & 8.79\% & 99.47\%\\  [1ex]
		\hline 
	\end{tabular}
\end{table}

It can be observed from the Table \ref{table:res} that VGG16 and Xception achieved state-of-the art results on the balanced dataset for pneumonia detection. Result of ResNet50 needs further investigation and can be improved with extensive hyper-parameter tuning and training with more epochs. 


\subsection{Comparison with the state of the art methods}

Table \ref{table:com} shows comparison of pneumonia detection accuracy of our proposed novel framework with state-of-the-art models. Although our proposed model achieved accuracy that exceeds state-of-the-art, directly comparing accuracy is not objective. This is due to the fact that models listed in Table \ref{table:com} are tested by respective researchers under different conditions / parameters. One such parameter is the variability in the dataset.

\begin{table}[!htb]
	
		\caption{Comparison of proposed approach with the state-of-the-art methods} 
	\label{table:com}

	\begin{tabular}{l|l|l|l|l}
		\hline
		\hline
		\textbf{Reference}                              & \textbf{Year} & \textbf{Method}                  & \textbf{Data}     & \textbf{Accuracy}  \\
		\hline
		Ozturk et al. \cite{Ozturk2020} 	   & 2020 & Yolo - DarkNet          & 1000 CXR & 87\%      \\
		Alhudjaif et al. \cite{ALHUDHAIF2021115141}   & 2021 & DenseNet-201            & 1218 CXR & 94.96\%   \\
		Das et al. \cite{Das2021}                     & 2021 & CNN + transfer learning & 1006 CXR & 91.62\%    \\
	Srivastav et al. \cite{Srivastav2021}  &2021   &DCGAN, CNN, TL &5856 CXR &94.5\% \\
	Nikolaou et al. \cite{Nikolaou2021} &2021 &EfficientNetB0, TL  &15,153 CXR & 95\%\\ 
		Singh and Tripathi \cite{Singh2022}           & 2022 & Quaternion CNN          & 5856 CXR & 93.75\%    \\
		Gour and Jain \cite{GOUR202227}               & 2022 & VGG-19, Xception        & 3040 CXR & 97.62\%    \\ &&&& (Sensitivity)        \\
		Szepesi and Szilagy \cite{SZEPESI2022}                   & 2022 & CNN + modified dropout  & 5856 CXR & 97.2\%        \\

		\textbf{Ours}  & \textbf{2023} & \textbf{DCGAN with Wasserstein GP, TL} &\textbf{60,000 CXR} &\textbf{89-99.4\%}\\
		
		\hline
	\end{tabular}
\end{table}

Apart from comparing accuracy it is also important to consider size of different models or number of parameters required to train various CNN models. ResNet50 has around 25 million parameters, VGG16 has 138 million parameters and Xception has  27 million parameters \cite{hassan2021identification}. With these statistics it is evident that different models will take different time to train and time for inference will also be different for a single example. Among the above-mentioned three models VGG16 has the highest number of trainable parameters.

\section{Conclusion and Future Work} \label{conc}

Pneumonia is one of the biggest threats to human life all over the world. Early diagnosis of pneumonia is essential to choose the best treatment method and further prevent infection that is endangering the patient's life. An X-Ray scan is frequently used to aid in the diagnosis of pneumonia. Regardless of the presence of pneumonia on the X-Ray images, the diagnosis is always dependent on the doctor's expertise and experience. However, due to a scarcity of competent medical radiologists in developing nations, examining chest X-Rays is a difficult undertaking as it is vulnerable to subjective uncertainty. There's no denying that an automated system is constantly needed to quicken the image analysis process and assist radiologists in diagnosing the deadly Pneumonia disease.

In this article, we propose an automated Computer-Aided Diagnosis (CAD) system to aid medical practitioners in the diagnosis of Pneumonia. We used ``Chest X-Ray8'' dataset \cite{wang2017chestx}, one of the most widely used and largest publicly available dataset related to lung diseases. The dataset distribution is severely skewed. To deal with the class imbalance problem we propose to use Generative Adversarial Network (GAN) \cite{goodfellow2014generative}, specifically a combination of Deep Convolutional Generative Adversarial Network (DCGAN) \cite{radford2015unsupervised} and Wasserstein GAN gradient penalty (WGAN-GP) \cite{gulrajani2017improved}. The proposed method generates sample for minority class i.e. ``Pneumonia''. Whereas Random Under-Sampling (RUS) \cite{khushi2021comparative} was applied on the majority class i.e. ``no-findings''. The balanced data is then used for training three state-of-the-art CNN models using the transfer learning paradigm. Proposed method achieved accuracy that exceeded state-of-the-art.

Some of the the limitations of our study are: 

\begin{enumerate}
	\item We relied on data from single source i.e. ``Chest X-Ray8'' dataset. Although it is one of the largest publicly available dataset, it still lacks diversity in X-ray capturing setup and human subjects. 
	\item The proposed system is only tested on Chest X-ray (CXR) images. It is also required to validate it for different imaging modalities e.g. computerized tomography (CT) scans, Magnetic Resonance Imaging (MRI) etc. 
	\item There is a need to verify that the AI / ML model is learning relevant features  for prediction. There is a need to induce explainability in prediction process or to use explainable AI models \cite{CSISZAR2020}. 
\end{enumerate}

Furthermore, we look forward to exploring additional classification techniques which may lead to development of our own CNN model instead of using a transfer learning approach. Additionally, we intend to explore the ensemble framework that is training a variety of classifiers and clusters altogether for producing better results instead of training a single classifier.




\begin{thebibliography}{10}
	\expandafter\ifx\csname url\endcsname\relax
	\def\url#1{\texttt{#1}}\fi
	\expandafter\ifx\csname urlprefix\endcsname\relax\def\urlprefix{URL }\fi
	\expandafter\ifx\csname href\endcsname\relax
	\def\href#1#2{#2} \def\path#1{#1}\fi
	
	\bibitem{ZH2}
	X.~Zhao, B.~Liu, Y.~Yu, X.~Wang, Y.~Du, J.~Gu, X.~Wu,
	\href{https://www.sciencedirect.com/science/article/pii/S0009926020300866}{The
		characteristics and clinical value of chest ct images of novel coronavirus
		pneumonia}, Clinical Radiology 75~(5) (2020) 335--340.
	\newblock \href {https://doi.org/https://doi.org/10.1016/j.crad.2020.03.002}
	{\path{doi:https://doi.org/10.1016/j.crad.2020.03.002}}.
	\newline\urlprefix\url{https://www.sciencedirect.com/science/article/pii/S0009926020300866}
	
	\bibitem{SMITH20}
	K.~R. Smith, J.~P. McCracken, M.~W. Weber, A.~Hubbard, A.~Jenny, L.~M.
	Thompson, J.~Balmes, A.~Diaz, B.~Arana, N.~Bruce,
	\href{https://www.sciencedirect.com/science/article/pii/S0140673611609215}{Effect
		of reduction in household air pollution on childhood pneumonia in guatemala
		{(RESPIRE)}: a randomised controlled trial}, The Lancet 378~(9804) (2011)
	1717--1726.
	\newblock \href {https://doi.org/https://doi.org/10.1016/S0140-6736(11)60921-5}
	{\path{doi:https://doi.org/10.1016/S0140-6736(11)60921-5}}.
	\newline\urlprefix\url{https://www.sciencedirect.com/science/article/pii/S0140673611609215}
	
	\bibitem{HARRIOTT201}
	M.~M. Harriott, M.~C. Noverr,
	\href{https://www.sciencedirect.com/science/article/pii/S0966842X11001417}{Importance
		of candida–bacterial polymicrobial biofilms in disease}, Trends in
	Microbiology 19~(11) (2011) 557--563.
	\newblock \href {https://doi.org/https://doi.org/10.1016/j.tim.2011.07.004}
	{\path{doi:https://doi.org/10.1016/j.tim.2011.07.004}}.
	\newline\urlprefix\url{https://www.sciencedirect.com/science/article/pii/S0966842X11001417}
	
	\bibitem{Pneumonia2021}
	Roa fact sheets, 2021 (2021).
	\newblock \href {https://doi.org/10.26481/umarof-2021}
	{\path{doi:10.26481/umarof-2021}}.
	
	\bibitem{owidpneumonia}
	B.~Dadonaite, M.~Roser, Pneumonia, Our World in
	DataHttps://ourworldindata.org/pneumonia (2018).
	
	\bibitem{Kondo2017}
	K.~Kondo, K.~Suzuki, M.~Washio, S.~Ohfuji, W.~Fukushima, A.~Maeda, Y.~Hirota,
	Effectiveness of 23-valent pneumococcal polysaccharide vaccine and seasonal
	influenza vaccine for pneumonia among the elderly – selection of controls
	in a case-control study, Vaccine 35 (2017) 4806--4810.
	\newblock \href {https://doi.org/10.1016/j.vaccine.2017.07.005}
	{\path{doi:10.1016/j.vaccine.2017.07.005}}.
	
	\bibitem{Doi2007}
	K.~Doi,
	\href{https://www.sciencedirect.com/science/article/pii/S0895611107000262}{Computer-aided
		diagnosis in medical imaging: Historical review, current status and future
		potential}, Computerized Medical Imaging and Graphics 31~(4) (2007) 198--211,
	computer-aided Diagnosis (CAD) and Image-guided Decision Support.
	\newblock \href
	{https://doi.org/https://doi.org/10.1016/j.compmedimag.2007.02.002}
	{\path{doi:https://doi.org/10.1016/j.compmedimag.2007.02.002}}.
	\newline\urlprefix\url{https://www.sciencedirect.com/science/article/pii/S0895611107000262}
	
	\bibitem{SHAH20221}
	S.~M. Shah, R.~A. Khan, S.~Arif, U.~Sajid,
	\href{https://www.sciencedirect.com/science/article/pii/S0010482522000130}{Artificial
		intelligence for breast cancer analysis: Trends \& directions}, Computers in
	Biology and Medicine 142 (2022) 105221.
	\newblock \href
	{https://doi.org/https://doi.org/10.1016/j.compbiomed.2022.105221}
	{\path{doi:https://doi.org/10.1016/j.compbiomed.2022.105221}}.
	\newline\urlprefix\url{https://www.sciencedirect.com/science/article/pii/S0010482522000130}
	
	\bibitem{Yu2018}
	K.-H. Yu, A.~L. Beam, I.~S. Kohane,
	\href{https://doi.org/10.1038/s41551-018-0305-z}{Artificial intelligence in
		healthcare}, Nature Biomedical Engineering 2~(10) (2018) 719--731.
	\newblock \href {https://doi.org/10.1038/s41551-018-0305-z}
	{\path{doi:10.1038/s41551-018-0305-z}}.
	\newline\urlprefix\url{https://doi.org/10.1038/s41551-018-0305-z}
	
	\bibitem{BAGCI201272}
	U.~Bağcı, M.~Bray, J.~Caban, J.~Yao, D.~J. Mollura,
	\href{https://www.sciencedirect.com/science/article/pii/S0895611111000802}{Computer-assisted
		detection of infectious lung diseases: A review}, Computerized Medical
	Imaging and Graphics 36~(1) (2012) 72--84.
	\newblock \href
	{https://doi.org/https://doi.org/10.1016/j.compmedimag.2011.06.002}
	{\path{doi:https://doi.org/10.1016/j.compmedimag.2011.06.002}}.
	\newline\urlprefix\url{https://www.sciencedirect.com/science/article/pii/S0895611111000802}
	
	\bibitem{Nishio2020}
	M.~Nishio, S.~Noguchi, H.~Matsuo, T.~Murakami,
	\href{https://doi.org/10.1038/s41598-020-74539-2}{Automatic classification
		between {COVID-19} pneumonia, non-{COVID-19} pneumonia, and the healthy on
		chest x-ray image: combination of data augmentation methods}, Scientific
	Reports 10~(1) (2020) 17532.
	\newblock \href {https://doi.org/10.1038/s41598-020-74539-2}
	{\path{doi:10.1038/s41598-020-74539-2}}.
	\newline\urlprefix\url{https://doi.org/10.1038/s41598-020-74539-2}
	
	\bibitem{gulati2021lung}
	A.~Gulati, R.~Balasubramanya, Lung imaging, in: StatPearls [Internet],
	StatPearls Publishing, 2021.
	
	\bibitem{9794709}
	M.~Yaseliani, A.~Z. Hamadani, A.~I. Maghsoodi, A.~Mosavi, Pneumonia detection
	proposing a hybrid deep convolutional neural network based on two parallel
	visual geometry group architectures and machine learning classifiers, IEEE
	Access 10 (2022) 62110--62128.
	\newblock \href {https://doi.org/10.1109/ACCESS.2022.3182498}
	{\path{doi:10.1109/ACCESS.2022.3182498}}.
	
	\bibitem{hwang2021covid}
	E.~J. Hwang, K.~B. Kim, J.~Y. Kim, J.-K. Lim, J.~G. Nam, H.~Choi, H.~Kim, S.~H.
	Yoon, J.~M. Goo, C.~M. Park, {COVID-19} pneumonia on chest {X-rays}:
	Performance of a deep learning-based computer-aided detection system, PLoS
	One 16~(6) (2021) e0252440.
	
	\bibitem{WANG2021107613}
	Z.~Wang, Y.~Xiao, Y.~Li, J.~Zhang, F.~Lu, M.~Hou, X.~Liu,
	\href{https://www.sciencedirect.com/science/article/pii/S0031320320304167}{Automatically
		discriminating and localizing {COVID-19} from community-acquired pneumonia on
		chest x-rays}, Pattern Recognition 110 (2021) 107613.
	\newblock \href {https://doi.org/https://doi.org/10.1016/j.patcog.2020.107613}
	{\path{doi:https://doi.org/10.1016/j.patcog.2020.107613}}.
	\newline\urlprefix\url{https://www.sciencedirect.com/science/article/pii/S0031320320304167}
	
	\bibitem{LeCun2015}
	Y.~LeCun, Y.~Bengio, G.~Hinton, Deep learning, Nature 521 (2015) 436--44.
	\newblock \href {https://doi.org/10.1038/nature14539}
	{\path{doi:10.1038/nature14539}}.
	
	\bibitem{Zhang2021}
	F.~Zhang,
	\href{https://www.ncbi.nlm.nih.gov/pmc/articles/PMC8428739/}{Application of
		machine learning in ct images and x-rays of covid-19 pneumonia}, Medicine
	100~(34516488) (2021) e26855--e26855.
	\newblock \href {https://doi.org/10.1097/MD.0000000000026855}
	{\path{doi:10.1097/MD.0000000000026855}}.
	\newline\urlprefix\url{https://www.ncbi.nlm.nih.gov/pmc/articles/PMC8428739/}
	
	\bibitem{HMRZ}
	H.~Sharif, R.~A. Khan, \href{https://doi.org/10.1080/08839514.2021.2004655}{A
		novel machine learning based framework for detection of autism spectrum
		disorder {(ASD)}}, Applied Artificial Intelligence 36~(1) (2022).
	\newblock \href
	{http://arxiv.org/abs/https://doi.org/10.1080/08839514.2021.2004655}
	{\path{arXiv:https://doi.org/10.1080/08839514.2021.2004655}}, \href
	{https://doi.org/10.1080/08839514.2021.2004655}
	{\path{doi:10.1080/08839514.2021.2004655}}.
	\newline\urlprefix\url{https://doi.org/10.1080/08839514.2021.2004655}
	
	\bibitem{LDong18}
	D.~Liang, L.~Lin, H.~Hu, Q.~Zhang, Q.~Chen, Y.~lwamoto, X.~Han, Y.-W. Chen,
	Combining convolutional and recurrent neural networks for classification of
	focal liver lesions in multi-phase {CT} images, in: A.~F. Frangi, J.~A.
	Schnabel, C.~Davatzikos, C.~Alberola-L{\'o}pez, G.~Fichtinger (Eds.), Medical
	Image Computing and Computer Assisted Intervention -- MICCAI 2018, Springer
	International Publishing, Cham, 2018, pp. 666--675.
	
	\bibitem{Gao2020}
	L.~Gao, L.~Zhang, C.~Liu, S.~Wu, Handling imbalanced medical image data: A
	deep-learning-based one-class classification approach., Artificial
	intelligence in medicine 108 (2020) 101935.
	\newblock \href {https://doi.org/10.1016/j.artmed.2020.101935}
	{\path{doi:10.1016/j.artmed.2020.101935}}.
	
	\bibitem{wang2017chestx}
	X.~Wang, Y.~Peng, L.~Lu, Z.~Lu, M.~Bagheri, R.~M. Summers, Chest x-ray8:
	Hospital-scale chest {X-ray} database and benchmarks on weakly-supervised
	classification and localization of common thorax diseases, in: Proceedings of
	the IEEE conference on computer vision and pattern recognition, 2017, pp.
	2097--2106.
	
	\bibitem{KHAN201961}
	R.~A. Khan, A.~Crenn, A.~Meyer, S.~Bouakaz,
	\href{https://www.sciencedirect.com/science/article/pii/S0262885619300137}{A
		novel database of children's spontaneous facial expressions {(LIRIS-CSE)}},
	Image and Vision Computing 83-84 (2019) 61--69.
	\newblock \href {https://doi.org/https://doi.org/10.1016/j.imavis.2019.02.004}
	{\path{doi:https://doi.org/10.1016/j.imavis.2019.02.004}}.
	\newline\urlprefix\url{https://www.sciencedirect.com/science/article/pii/S0262885619300137}
	
	\bibitem{STcomplex}
	E.~C. Alfredo~Canziani, Adam~Paszke, An analysis of deep neural network models
	for practical applications, arXiv:1605.07678 (2017).
	
	\bibitem{goodfellow2014generative}
	I.~Goodfellow, J.~Pouget-Abadie, M.~Mirza, B.~Xu, D.~Warde-Farley, S.~Ozair,
	A.~Courville, Y.~Bengio, Generative adversarial nets, Advances in neural
	information processing systems 27 (2014).
	\newblock \href {https://doi.org/10.1145/3422622} {\path{doi:10.1145/3422622}}.
	
	\bibitem{radford2015unsupervised}
	A.~Radford, L.~Metz, S.~Chintala, Unsupervised representation learning with
	deep convolutional generative adversarial networks, arXiv preprint
	arXiv:1511.06434 (2015).
	\newblock \href {http://arxiv.org/abs/1511.06434} {\path{arXiv:1511.06434}}.
	
	\bibitem{gulrajani2017improved}
	I.~Gulrajani, F.~Ahmed, M.~Arjovsky, V.~Dumoulin, A.~C. Courville, Improved
	training of wasserstein gans, Advances in neural information processing
	systems 30 (2017).
	\newblock \href {http://arxiv.org/abs/1704.00028} {\path{arXiv:1704.00028}}.
	
	\bibitem{Davenport2020}
	T.~Davenport, A.~Guha, D.~Grewal, T.~Bressgott,
	\href{https://doi.org/10.1007/s11747-019-00696-0}{How artificial intelligence
		will change the future of marketing}, Journal of the Academy of Marketing
	Science 48~(1) (2020) 24--42.
	\newblock \href {https://doi.org/10.1007/s11747-019-00696-0}
	{\path{doi:10.1007/s11747-019-00696-0}}.
	\newline\urlprefix\url{https://doi.org/10.1007/s11747-019-00696-0}
	
	\bibitem{4490127}
	J.~D. Owens, M.~Houston, D.~Luebke, S.~Green, J.~E. Stone, J.~C. Phillips, {GPU
		Computing}, Proceedings of the IEEE 96~(5) (2008) 879--899.
	\newblock \href {https://doi.org/10.1109/JPROC.2008.917757}
	{\path{doi:10.1109/JPROC.2008.917757}}.
	
	\bibitem{Hinton2006}
	G.~E. Hinton, R.~R. Salakhutdinov, Reducing the dimensionality of data with
	neural networks., Science (New York, N.Y.) 313 (2006) 504--7.
	\newblock \href {https://doi.org/10.1126/science.1127647}
	{\path{doi:10.1126/science.1127647}}.
	
	\bibitem{Srivastava2014}
	N.~Srivastava, G.~Hinton, A.~Krizhevsky, I.~Sutskever, R.~Salakhutdinov,
	Dropout: A simple way to prevent neural networks from overfitting, Journal of
	Machine Learning Research 15 (2014) 1929--1958.
	
	\bibitem{Girshick2014}
	R.~B. Girshick, J.~Donahue, T.~Darrell, J.~Malik, Rich feature hierarchies for
	accurate object detection and semantic segmentation, 2014 IEEE Conference on
	Computer Vision and Pattern Recognition (2014) 580--587\href
	{https://doi.org/10.1109/cvpr.2014.81} {\path{doi:10.1109/cvpr.2014.81}}.
	
	\bibitem{Voulodimos2018}
	A.~Voulodimos, N.~Doulamis, A.~Doulamis, E.~Protopapadakis,
	\href{https://doi.org/10.1155/2018/7068349}{Deep learning for computer
		vision: A brief review}, Computational Intelligence and Neuroscience 2018
	(2018) 7068349.
	\newblock \href {https://doi.org/10.1155/2018/7068349}
	{\path{doi:10.1155/2018/7068349}}.
	\newline\urlprefix\url{https://doi.org/10.1155/2018/7068349}
	
	\bibitem{lecun1998gradient}
	Y.~LeCun, L.~Bottou, Y.~Bengio, P.~Haffner, Gradient-based learning applied to
	document recognition, Proceedings of the IEEE 86~(11) (1998) 2278--2324.
	\newblock \href {https://doi.org/10.1109/5.726791}
	{\path{doi:10.1109/5.726791}}.
	
	\bibitem{Fukushima1980}
	K.~Fukushima, \href{https://doi.org/10.1007/BF00344251}{Neocognitron: A
		self-organizing neural network model for a mechanism of pattern recognition
		unaffected by shift in position}, Biological Cybernetics 36~(4) (1980)
	193--202.
	\newblock \href {https://doi.org/10.1007/BF00344251}
	{\path{doi:10.1007/BF00344251}}.
	\newline\urlprefix\url{https://doi.org/10.1007/BF00344251}
	
	\bibitem{arxiv17110}
	P.~Rajpurkar, J.~Irvin, K.~Zhu, B.~Yang, H.~Mehta, T.~Duan, D.~Ding, A.~Bagul,
	C.~Langlotz, K.~Shpanskaya, M.~P. Lungren, A.~Y. Ng,
	\href{https://arxiv.org/abs/1711.05225}{{CheXNet}: Radiologist-level
		pneumonia detection on chest {X-Rays} with deep learning} (2017).
	\newblock \href {https://doi.org/10.48550/ARXIV.1711.05225}
	{\path{doi:10.48550/ARXIV.1711.05225}}.
	\newline\urlprefix\url{https://arxiv.org/abs/1711.05225}
	
	\bibitem{Aplos}
	J.~M. Andrew G~Taylor, Clinton~Mielke, Automated detection of moderate and
	large pneumothorax on frontal chest x-rays using deep convolutional neural
	networks: A retrospective study, PLoS medicine (2018).
	
	\bibitem{EKANEM2021437}
	E.~Ekanem, S.~Podder, N.~Donthi, H.~Bakhshi, J.~Stodghill, S.~Khandhar,
	A.~Mahajan, M.~Desai,
	\href{https://www.sciencedirect.com/science/article/pii/S0147956321000200}{Spontaneous
		pneumothorax: An emerging complication of {COVID-19} pneumonia}, Heart \&
	Lung 50~(3) (2021) 437--440.
	\newblock \href {https://doi.org/https://doi.org/10.1016/j.hrtlng.2021.01.020}
	{\path{doi:https://doi.org/10.1016/j.hrtlng.2021.01.020}}.
	\newline\urlprefix\url{https://www.sciencedirect.com/science/article/pii/S0147956321000200}
	
	\bibitem{Ozturk2020}
	T.~Ozturk, M.~Talo, E.~A. Yildirim, U.~B. Baloglu, O.~Yildirim, U.~{Rajendra
		Acharya},
	\href{https://www.sciencedirect.com/science/article/pii/S0010482520301621}{Automated
		detection of covid-19 cases using deep neural networks with x-ray images},
	Computers in Biology and Medicine 121 (2020) 103792.
	\newblock \href
	{https://doi.org/https://doi.org/10.1016/j.compbiomed.2020.103792}
	{\path{doi:https://doi.org/10.1016/j.compbiomed.2020.103792}}.
	\newline\urlprefix\url{https://www.sciencedirect.com/science/article/pii/S0010482520301621}
	
	\bibitem{AlMamlook2020}
	R.~Al~Mamlook, S.~Chen, H.~Bzizi, Investigation of the performance of machine
	learning classifiers for pneumonia detection in chest {X-ray} images, in:
	IEEE International Conference on Electro Information Technology (EIT), 2020.
	\newblock \href {https://doi.org/10.1109/EIT48999.2020.9208232}
	{\path{doi:10.1109/EIT48999.2020.9208232}}.
	
	\bibitem{KHAN20131159}
	R.~A. Khan, A.~Meyer, H.~Konik, S.~Bouakaz,
	\href{https://www.sciencedirect.com/science/article/pii/S0167865513001268}{Framework
		for reliable, real-time facial expression recognition for low resolution
		images}, Pattern Recognition Letters 34~(10) (2013) 1159--1168.
	\newblock \href {https://doi.org/https://doi.org/10.1016/j.patrec.2013.03.022}
	{\path{doi:https://doi.org/10.1016/j.patrec.2013.03.022}}.
	\newline\urlprefix\url{https://www.sciencedirect.com/science/article/pii/S0167865513001268}
	
	\bibitem{ALHUDHAIF2021115141}
	A.~Alhudhaif, K.~Polat, O.~Karaman,
	\href{https://www.sciencedirect.com/science/article/pii/S0957417421005820}{Determination
		of {COVID-19} pneumonia based on generalized convolutional neural network
		model from chest {X-ray} images}, Expert Systems with Applications 180 (2021)
	115141.
	\newblock \href {https://doi.org/https://doi.org/10.1016/j.eswa.2021.115141}
	{\path{doi:https://doi.org/10.1016/j.eswa.2021.115141}}.
	\newline\urlprefix\url{https://www.sciencedirect.com/science/article/pii/S0957417421005820}
	
	\bibitem{Huang2017}
	G.~Huang, Z.~Liu, L.~van~der Maaten, K.~Weinberger, Densely connected
	convolutional networks, in: IEEE Conference on Computer Vision and Pattern
	Recognition (CVPR), 2017.
	\newblock \href {https://doi.org/10.1109/CVPR.2017.243}
	{\path{doi:10.1109/CVPR.2017.243}}.
	
	\bibitem{Nikolaou2021}
	V.~Nikolaou, S.~Massaro, M.~Fakhimi, L.~Stergioulas, W.~Garn,
	\href{https://doi.org/10.1007/s13755-021-00166-4}{{COVID-19} diagnosis from
		chest {X-rays}: developing a simple, fast, and accurate neural network},
	Health Information Science and Systems 9~(1) (2021) 36.
	\newblock \href {https://doi.org/10.1007/s13755-021-00166-4}
	{\path{doi:10.1007/s13755-021-00166-4}}.
	\newline\urlprefix\url{https://doi.org/10.1007/s13755-021-00166-4}
	
	\bibitem{DatabaseKag}
	Covid-19 radiography database,
	https://www.kaggle.com/datasets/tawsifurrahman/covid19-radiography-database.
	
	\bibitem{pmlr-v97-tan19a}
	M.~Tan, Q.~Le,
	\href{https://proceedings.mlr.press/v97/tan19a.html}{{E}fficient{N}et:
		Rethinking model scaling for convolutional neural networks}, in:
	K.~Chaudhuri, R.~Salakhutdinov (Eds.), Proceedings of the 36th International
	Conference on Machine Learning, Vol.~97 of Proceedings of Machine Learning
	Research, PMLR, 2019, pp. 6105--6114.
	\newline\urlprefix\url{https://proceedings.mlr.press/v97/tan19a.html}
	
	\bibitem{Das2021}
	A.~K. Das, S.~Ghosh, S.~Thunder, R.~Dutta, S.~Agarwal, A.~Chakrabarti,
	\href{https://doi.org/10.1007/s10044-021-00970-4}{Automatic {COVID-19}
		detection from {X-ray} images using ensemble learning with convolutional
		neural network}, Pattern Analysis and Applications 24~(3) (2021) 1111--1124.
	\newblock \href {https://doi.org/10.1007/s10044-021-00970-4}
	{\path{doi:10.1007/s10044-021-00970-4}}.
	\newline\urlprefix\url{https://doi.org/10.1007/s10044-021-00970-4}
	
	\bibitem{He2016}
	K.~He, X.~Zhang, S.~Ren, J.~Sun, Identity mappings in deep residual networks,
	in: B.~Leibe, J.~Matas, N.~Sebe, M.~Welling (Eds.), Computer Vision -- ECCV
	2016, Springer International Publishing, Cham, 2016, pp. 630--645.
	\newblock \href {https://doi.org/10.1007/978-3-319-46493-0_38}
	{\path{doi:10.1007/978-3-319-46493-0_38}}.
	
	\bibitem{szegedy2015going}
	C.~Szegedy, W.~Liu, Y.~Jia, P.~Sermanet, S.~Reed, D.~Anguelov, D.~Erhan,
	V.~Vanhoucke, A.~Rabinovich, Going deeper with convolutions, in: Proceedings
	of the IEEE conference on computer vision and pattern recognition, 2015, pp.
	1--9.
	\newblock \href {http://arxiv.org/abs/1409.4842} {\path{arXiv:1409.4842}}.
	
	\bibitem{Singh2022}
	S.~Singh, B.~K. Tripathi,
	\href{https://doi.org/10.1007/s11042-021-11409-7}{Pneumonia classification
		using quaternion deep learning}, Multimedia Tools and Applications 81~(2)
	(2022) 1743--1764.
	\newblock \href {https://doi.org/10.1007/s11042-021-11409-7}
	{\path{doi:10.1007/s11042-021-11409-7}}.
	\newline\urlprefix\url{https://doi.org/10.1007/s11042-021-11409-7}
	
	\bibitem{Zhu_2018_ECCV}
	X.~Zhu, Y.~Xu, H.~Xu, C.~Chen, Quaternion convolutional neural networks, in:
	Proceedings of the European Conference on Computer Vision (ECCV), 2018.
	
	\bibitem{GOUR202227}
	M.~Gour, S.~Jain,
	\href{https://www.sciencedirect.com/science/article/pii/S0208521621001303}{Automated
		{COVID-19} detection from {X-ray} and {CT} images with stacked ensemble
		convolutional neural network}, Biocybernetics and Biomedical Engineering
	42~(1) (2022) 27--41.
	\newblock \href {https://doi.org/https://doi.org/10.1016/j.bbe.2021.12.001}
	{\path{doi:https://doi.org/10.1016/j.bbe.2021.12.001}}.
	\newline\urlprefix\url{https://www.sciencedirect.com/science/article/pii/S0208521621001303}
	
	\bibitem{Simonyan2015}
	K.~Simonyan, A.~Zisserman, Very deep convolutional networks for large-scale
	image recognition, CoRR:10.48550/arXiv.1409.1556 (2015).
	\newblock \href {http://arxiv.org/abs/1409.1556} {\path{arXiv:1409.1556}}.
	
	\bibitem{chollet2017xception}
	F.~Chollet, Xception: Deep learning with depthwise separable convolutions, in:
	Proceedings of the IEEE conference on computer vision and pattern
	recognition, 2017, pp. 1251--1258.
	
	\bibitem{SZEPESI2022}
	P.~Szepesi, L.~Szilágyi,
	\href{https://www.sciencedirect.com/science/article/pii/S0208521622000742}{Detection
		of pneumonia using convolutional neural networks and deep learning},
	Biocybernetics and Biomedical Engineering 42~(3) (2022) 1012--1022.
	\newblock \href {https://doi.org/https://doi.org/10.1016/j.bbe.2022.08.001}
	{\path{doi:https://doi.org/10.1016/j.bbe.2022.08.001}}.
	\newline\urlprefix\url{https://www.sciencedirect.com/science/article/pii/S0208521622000742}
	
	\bibitem{GoodfellowB2016}
	I.~Goodfellow, Y.~Bengio, A.~Courville, Deep Learning, MIT Press, 2016,
	\url{http://www.deeplearningbook.org}.
	
	\bibitem{mariani2018bagan}
	G.~Mariani, F.~Scheidegger, R.~Istrate, C.~Bekas, C.~Malossi, {BAGAN}: Data
	augmentation with balancing {GAN}, arXiv preprint arXiv:1803.09655 (2018).
	
	\bibitem{Borjali2020}
	A.~Borjali, A.~F. Chen, O.~K. Muratoglu, M.~A. Morid, K.~M. Varadarajan,
	Detecting total hip replacement prosthesis design on plain radiographs using
	deep convolutional neural network., Journal of orthopaedic research :
	official publication of the Orthopaedic Research Society 38 (2020)
	1465--1471.
	\newblock \href {https://doi.org/10.1002/jor.24617}
	{\path{doi:10.1002/jor.24617}}.
	
	\bibitem{Leevy2018}
	J.~L. Leevy, T.~M. Khoshgoftaar, R.~A. Bauder, N.~Seliya,
	\href{https://doi.org/10.1186/s40537-018-0151-6}{A survey on addressing
		high-class imbalance in big data}, Journal of Big Data 5~(1) (2018) 42.
	\newblock \href {https://doi.org/10.1186/s40537-018-0151-6}
	{\path{doi:10.1186/s40537-018-0151-6}}.
	\newline\urlprefix\url{https://doi.org/10.1186/s40537-018-0151-6}
	
	\bibitem{roughgarden2010algorithmic}
	T.~Roughgarden, Algorithmic game theory, Communications of the ACM 53~(7)
	(2010) 78--86.
	
	\bibitem{wang2017irgan}
	J.~Wang, L.~Yu, W.~Zhang, Y.~Gong, Y.~Xu, B.~Wang, P.~Zhang, D.~Zhang, {IRGAN}:
	A minimax game for unifying generative and discriminative information
	retrieval models, in: Proceedings of the 40th International ACM SIGIR
	conference on Research and Development in Information Retrieval, 2017, pp.
	515--524.
	
	\bibitem{irvin2019chexpert}
	J.~Irvin, P.~Rajpurkar, M.~Ko, Y.~Yu, S.~Ciurea-Ilcus, C.~Chute, H.~Marklund,
	B.~Haghgoo, R.~Ball, K.~Shpanskaya, et~al., {Chexpert:} a large chest
	radiograph dataset with uncertainty labels and expert comparison, in:
	Proceedings of the {AAAI} conference on artificial intelligence, Vol.~33,
	2019, pp. 590--597.
	
	\bibitem{Sundaram2021}
	S.~Sundaram, N.~Hulkund, {GAN-}based data augmentation for {Chest X-ray}
	classification, ArXiv abs/2107.02970 (2021).
	\newblock \href {http://arxiv.org/abs/2107.02970} {\path{arXiv:2107.02970}}.
	
	\bibitem{Malygina2019}
	T.~Malygina, E.~Ericheva, I.~Drokin, Data augmentation with {GAN:} improving
	chest x-ray pathologies prediction on class-imbalanced cases, in: W.~M.~P.
	van~der Aalst, V.~Batagelj, D.~I. Ignatov, M.~Khachay, V.~Kuskova,
	A.~Kutuzov, S.~O. Kuznetsov, I.~A. Lomazova, N.~Loukachevitch, A.~Napoli,
	P.~M. Pardalos, M.~Pelillo, A.~V. Savchenko, E.~Tutubalina (Eds.), Analysis
	of Images, Social Networks and Texts, Springer International Publishing,
	Cham, 2019, pp. 321--334.
	\newblock \href {https://doi.org/10.1007/978-3-030-37334-4_29}
	{\path{doi:10.1007/978-3-030-37334-4_29}}.
	
	\bibitem{Srivastav2021}
	D.~Srivastav, A.~Bajpai, P.~Srivastava, Improved classification for pneumonia
	detection using transfer learning with {GAN} based synthetic image
	augmentation, in: 2021 11th International Conference on Cloud Computing, Data
	Science Engineering (Confluence), 2021, pp. 433--437.
	\newblock \href {https://doi.org/10.1109/Confluence51648.2021.9377062}
	{\path{doi:10.1109/Confluence51648.2021.9377062}}.
	
	\bibitem{kermany2018large}
	D.~Kermany, K.~Zhang, M.~Goldbaum, Large dataset of labeled optical coherence
	tomography {(OCT)} and chest x-ray images, Mendeley Data 3 (2018) 10--17632.
	
	\bibitem{sundaram2021gan}
	S.~Sundaram, N.~Hulkund, {GAN-based data augmentation for chest X-ray
		classification}, arXiv preprint arXiv:2107.02970 (2021).
	
	\bibitem{han2022gan}
	L.~Han, Y.~Lyu, C.~Peng, S.~K. Zhou, {GAN-based disentanglement learning for
		chest X-ray rib suppression}, Medical Image Analysis 77 (2022) 102369.
	
	\bibitem{shah2022dc}
	P.~M. Shah, H.~Ullah, R.~Ullah, D.~Shah, Y.~Wang, S.~u. Islam, A.~Gani, J.~J.
	Rodrigues, {DC-GAN-based synthetic X-ray images augmentation for increasing
		the performance of EfficientNet for COVID-19 detection}, Expert Systems
	39~(3) (2022) e12823.
	
	\bibitem{1056489}
	S.~Lloyd, {Least squares quantization in PCM}, {IEEE Transactions on
		Information Theory} 28~(2) (1982) 129--137.
	\newblock \href {https://doi.org/10.1109/TIT.1982.1056489}
	{\path{doi:10.1109/TIT.1982.1056489}}.
	
	\bibitem{mildenberger2002}
	P.~Mildenberger, M.~Eichelberg, E.~Martin, Introduction to the {DICOM}
	standard, European radiology 12~(4) (2002) 920--927.
	
	\bibitem{collobert2011natural}
	R.~Collobert, J.~Weston, L.~Bottou, M.~Karlen, K.~Kavukcuoglu, P.~Kuksa,
	{Natural language processing (almost) from scratch}, Journal of machine
	learning research 12~(ARTICLE) (2011) 2493--2537.
	
	\bibitem{leaman2015challenges}
	R.~Leaman, R.~Khare, Z.~Lu, Challenges in clinical natural language processing
	for automated disorder normalization, Journal of biomedical informatics 57
	(2015) 28--37.
	
	\bibitem{aronson2010overview}
	A.~R. Aronson, F.-M. Lang, An overview of {MetaMap}: historical perspective and
	recent advances, Journal of the American Medical Informatics Association
	17~(3) (2010) 229--236.
	
	\bibitem{khushi2021comparative}
	M.~Khushi, K.~Shaukat, T.~M. Alam, I.~A. Hameed, S.~Uddin, S.~Luo, X.~Yang,
	M.~C. Reyes, A comparative performance analysis of data resampling methods on
	imbalance medical data, {IEEE Access} 9 (2021) 109960--109975.
	
	\bibitem{hu2021exposing}
	S.~Hu, Y.~Li, S.~Lyu, Exposing gan-generated faces using inconsistent corneal
	specular highlights, in: ICASSP 2021-2021 IEEE International Conference on
	Acoustics, Speech and Signal Processing (ICASSP), IEEE, 2021, pp. 2500--2504.
	\newblock \href {https://doi.org/10.1109/icassp39728.2021.9414582}
	{\path{doi:10.1109/icassp39728.2021.9414582}}.
	
	\bibitem{Yang2019}
	X.~Yang, Y.~Li, H.~Qi, S.~Lyu,
	\href{https://doi.org/10.1145/3335203.3335724}{Exposing {GAN}-synthesized
		faces using landmark locations}, in: Proceedings of the ACM Workshop on
	Information Hiding and Multimedia Security, IH\&amp;MMSec'19, Association for
	Computing Machinery, New York, NY, USA, 2019, p. 113–118.
	\newblock \href {https://doi.org/10.1145/3335203.3335724}
	{\path{doi:10.1145/3335203.3335724}}.
	\newline\urlprefix\url{https://doi.org/10.1145/3335203.3335724}
	
	\bibitem{kazeminia2020gans}
	S.~Kazeminia, C.~Baur, A.~Kuijper, B.~van Ginneken, N.~Navab, S.~Albarqouni,
	A.~Mukhopadhyay, Gans for medical image analysis, Artificial Intelligence in
	Medicine 109 (2020) 101938.
	\newblock \href {https://doi.org/10.1016/j.artmed.2020.101938}
	{\path{doi:10.1016/j.artmed.2020.101938}}.
	
	\bibitem{hitawala2018comparative}
	S.~Hitawala, Comparative study on generative adversarial networks, arXiv
	preprint arXiv:1801.04271 (2018).
	
	\bibitem{sutter2020multimodal}
	T.~Sutter, I.~Daunhawer, J.~Vogt, Multimodal generative learning utilizing
	jensen-shannon-divergence, Advances in Neural Information Processing Systems
	33 (2020) 6100--6110.
	
	\bibitem{goodfellow2016nips}
	I.~Goodfellow, Nips 2016 tutorial: Generative adversarial networks, arXiv
	preprint arXiv:1701.00160 (2016).
	
	\bibitem{kingma2014adam}
	D.~P. Kingma, J.~Ba, {Adam: A method for stochastic optimization}, arXiv
	preprint arXiv:1412.6980 (2014).
	
	\bibitem{arjovsky2017wasserstein}
	M.~Arjovsky, S.~Chintala, L.~Bottou, Wasserstein generative adversarial
	networks, in: International conference on machine learning, PMLR, 2017, pp.
	214--223.
	
	\bibitem{salimans2016improved}
	T.~Salimans, I.~Goodfellow, W.~Zaremba, V.~Cheung, A.~Radford, X.~Chen,
	{Improved techniques for training GANs}, Advances in neural information
	processing systems 29 (2016).
	
	\bibitem{Pan2010}
	S.~J. Pan, Q.~Yang, A survey on transfer learning, IEEE Transactions on
	Knowledge and Data Engineering 22~(10) (2010) 1345--1359.
	\newblock \href {https://doi.org/10.1109/TKDE.2009.191}
	{\path{doi:10.1109/TKDE.2009.191}}.
	
	\bibitem{Liu1999}
	Y.~Liu, X.~Yao, Ensemble learning via negative correlation., Neural networks :
	the official journal of the International Neural Network Society 12 (1999)
	1399--1404.
	\newblock \href {https://doi.org/10.1016/s0893-6080(99)00073-8}
	{\path{doi:10.1016/s0893-6080(99)00073-8}}.
	
	\bibitem{Russakovsky2015}
	O.~Russakovsky, J.~Deng, H.~Su, J.~Krause, S.~Satheesh, S.~Ma, Z.~Huang,
	A.~Karpathy, A.~Khosla, M.~Bernstein, A.~C. Berg, L.~Fei-Fei,
	\href{https://doi.org/10.1007/s11263-015-0816-y}{Imagenet large scale visual
		recognition challenge}, International Journal of Computer Vision 115~(3)
	(2015) 211--252.
	\newblock \href {https://doi.org/10.1007/s11263-015-0816-y}
	{\path{doi:10.1007/s11263-015-0816-y}}.
	\newline\urlprefix\url{https://doi.org/10.1007/s11263-015-0816-y}
	
	\bibitem{Krizhevsky2012}
	A.~Krizhevsky, I.~Sutskever, G.~E. Hinton,
	\href{https://proceedings.neurips.cc/paper/2012/file/c399862d3b9d6b76c8436e924a68c45b-Paper.pdf}{Imagenet
		classification with deep convolutional neural networks}, in: F.~Pereira,
	C.~J.~C. Burges, L.~Bottou, K.~Q. Weinberger (Eds.), Advances in Neural
	Information Processing Systems, Vol.~25, Curran Associates, Inc., 2012.
	\newblock \href {https://doi.org/10.1145/3065386} {\path{doi:10.1145/3065386}}.
	\newline\urlprefix\url{https://proceedings.neurips.cc/paper/2012/file/c399862d3b9d6b76c8436e924a68c45b-Paper.pdf}
	
	\bibitem{He2015}
	K.~He, X.~Zhalng, S.~Ren, J.~Sun, Deep residual learning for image recognition
	(Dec. 2015).
	\newblock \href {http://arxiv.org/abs/1512.03385} {\path{arXiv:1512.03385}}.
	
	\bibitem{huang2017densely}
	G.~Huang, Z.~Liu, L.~Van Der~Maaten, K.~Q. Weinberger, Densely connected
	convolutional networks, in: IEEE conference on computer vision and pattern
	recognition, 2017, pp. 4700--4708.
	
	\bibitem{sajid2022breast}
	U.~Sajid, R.~Khan, M.~Shah, S.~Arif, Breast cancer classification using deep
	learned features boosted with handcrafted features, arXiv preprint
	arXiv:2206.12815 (2022).
	
	\bibitem{hassan2021identification}
	S.~M. Hassan, A.~K. Maji, M.~Jasi{\'n}ski, Z.~Leonowicz, E.~Jasi{\'n}ska,
	{Identification of plant-leaf diseases using CNN and transfer-learning
		approach}, Electronics 10~(12) (2021) 1388.
	
	\bibitem{CSISZAR2020}
	O.~Csiszár, G.~Csiszár, J.~Dombi,
	\href{https://www.sciencedirect.com/science/article/pii/S0950705120306596}{{How
			to implement MCDM tools and continuous logic into neural computation?:
			Towards better interpretability of neural networks}}, Knowledge-Based Systems
	210 (2020) 106530.
	\newblock \href {https://doi.org/https://doi.org/10.1016/j.knosys.2020.106530}
	{\path{doi:https://doi.org/10.1016/j.knosys.2020.106530}}.
	\newline\urlprefix\url{https://www.sciencedirect.com/science/article/pii/S0950705120306596}
	
\end{thebibliography}


\end{document}